# Dark-ring cavity solitons
# in lasers with bichromatic injected signal

**Germán J. de Valcárcel**
*Departament d'Òptica, Universitat de València, Dr. Moliner 50, 46100-Burjassot, Spain*
Telephone: +34 963 864 766 (ext. 16), fax: +34 963 864 715, e-mail address: german.valcarcel@uv.es

**Kestutis Staliunas**
*Physikalisch Technische Bundesanstalt, Bundesallee 100, 38116 Braunschweig, Germany*
Telephone:+49 531 592 4482, fax: +49 531 592 4423, e-mail address: kestutis.staliunas@ptb.de

**Abstract:** We show theoretically that broad area lasers driven by a nearly resonant bichromatic field may support dark-ring cavity solitons as well as domain walls and labyrinthine patterns.

**1. Introduction**

Many nonlinear optical cavities have been shown to support transverse patterns with different symmetries (rolls, hexagons, and also quasi-patterns) as well as cavity solitons (CS) [1]. The latter are particularly interesting objects because of their potential use as memory bits in optical information processing systems [2]. However usual lasers do not fall into the category of CS-supporting systems since, usually, either phase-sensitivity and/or a bistability is required for CS. Up to now two ways have been proposed for exciting CSs in lasers: the use of saturable absorbers [3], or the injection of a monochromatic signal [4]. Also very recently two-photon lasers have been predicted to support CS [5].

Here we demonstrate theoretically an alternative way for exciting CSs in lasers based on the injection of a resonant bichromatic signal into the laser cavity. Differently from the previously suggested laser CSs, the ones we present here are based on the development of two dynamically equivalent laser field phases differing by $\pi$. Hence the reported CSs belong to the class of dark-ring cavity solitons, in which a bright spot with a definite phase is surrounded by an oppositely phased, spatially uniform light distribution, both phases being separated by a dark ring. These phases appear when the laser emission frequency locks halfway between the two injected signal components, which can be understood as a result of the competition between a (phase sensitive) resonant, degenerate four-wave mixing process –in which two injected photons (one of each signal component) are converted into two equal laser photons– and the usual (phase insensitive) stimulated emission process of free running lasers.

Apart from CSs other phase patterns (e.g. domain walls or labyrinths) are shown to form in the laser transverse plane. We find two generic types of phase patterns, which differ in the way the light phase varies between two oppositely phased regions. In the first type the phase abruptly changes by $\pi$ along the line separating both regions; in these patterns the referred lines are dark and we speak of Ising-like patterns, in analogy with the Ising walls of ferromagnets [6]. In the second type the phase varies continuously and smoothly across the line, still accumulating $\pi$ between both sides; in these patterns the referred lines are gray and we speak of Bloch-like patterns [6].

As we show below the detuning between the central injection frequency and the free-running laser frequency selects the geometry of the observed patterns (domain walls, labyrinths, or dark-ring CSs), as well as the injected signal strength controls their Ising or Bloch character.

## 2. Model

We consider the standard Maxwell-Bloch equations [7] that model pattern formation in a single longitudinal mode, two-level laser with injected signal, inside a plane mirrors resonator, which can be written in dimensionless form as

$$\partial_t E = \sigma[-(1+i\Delta)E + P] + i\nabla^2 E + E_{\text{in}}, \qquad (1)$$

$$\partial_t P = -(1-i\Delta)P + (r-N)E, \qquad (2)$$

$$\partial_t N = b[-N + \text{Re}(E^*P)]. \qquad (3)$$

The complex fields $E$ and $P$ are the scaled envelopes of the electric field and medium polarization, $-N$ is proportional to the difference between the population inversion and its steady value in the absence of lasing, and $E_{\text{in}}$ is the scaled complex envelope of the injected signal. $\sigma = \kappa/\gamma_\perp$, and $b = \gamma_\parallel/\gamma_\perp$, being $\kappa = cT/L$ the cavity linewidth ($c$ is the effective light velocity within the resonator, $L$ is the cavity length, and $T$ is the mirrors' transmissivity), and $\gamma_\perp$, and $\gamma_\parallel$ the decay rates of $P$, and $N$, respectively. The transverse Laplacian $\nabla^2 = \partial_x^2 + \partial_y^2$, where the spatial coordinates $\mathbf{r} = (x, y)$ are measured in units of $(\gamma_\perp \omega_L)^{-1/2} c$, being $\omega_L$ the laser emission frequency; $t$ is time measured in units of $1/\gamma_\perp$. Finally $r$ is the usual laser pump parameter and the detuning $\Delta = (\omega_C - \omega_A)/(\gamma_\perp + \kappa)$, being $\omega_C (\omega_A)$ the cavity (atomic) frequency.

Eqs. (1)–(3) have been written in the frequency frame $\omega_0 = (\gamma_\perp \omega_C + \kappa \omega_A)/(\gamma_\perp + \kappa)$ of the on-axis, or plane wave $(\partial_x E = \partial_y E = 0)$, steady lasing solution in the absence of injected signal. Previous studies have considered pattern formation in model (1)–(3) when the injected signal is monochromatic, either as $E_{\text{in}} = E_0 \exp(-i\theta t)$ [7,8], or as $E_{\text{in}} = E_0 \exp(ikx - i\theta t)$ [9], with $E_0$ constant. These represent an on-axis, resp. a tilted, injected signal whose frequency detuning from $\omega_0$ is $\gamma_\perp \theta$. Here we consider the case

$$E_{\text{in}} = E_0 \cos(\omega t) \exp(-i\theta t), \qquad (4)$$

which represents an amplitude modulated field of carrier frequency $\omega_0 + \gamma_\perp \theta$, and modulation frequency $\gamma_\perp \omega$. Alternatively (4) represents a bichromatic injection formed by the superposition of two coherent light beams of frequencies $\omega_0 + \gamma_\perp \theta \pm \gamma_\perp \omega$, and equal amplitudes. In order that (4) be consistent with the uniform field and single longitudinal mode approximations that lead to Eqs. (1)–(3), both the carrier frequency offset $\gamma_\perp \theta$ and the modulation frequency $\gamma_\perp \omega$ must be much smaller than the cavity free spectral range $\alpha = 2\pi c/L = 2\pi \kappa/T$. Hence we must impose $|\theta|, |\omega| \ll 2\pi\sigma/T$. Note however that the rhs of the inequality is much larger than $\sigma$ since $T \ll 1$ is assumed in the uniform field limit, hence in fact there is almost no practical limitation to the values of $\theta$ and $\omega$.

In all this work we shall consider the positive detuning case $\Delta > 0$ for which, in the absence of injected signal, the laser is off for $r < r_0 \equiv 1 + \Delta^2$, and switches on at $r = r_0$ giving rise to the on-axis lasing solution [10,11].

## 3. Reduction of the Maxwell-Bloch equations to a driven, complex Ginzburg-Landau equation

In order to gain insight into the basic pattern forming properties of the system, we undergo next a reduction of Eqs. (1)–(3) to a compact form that allows a simplified treatment of the problem. The reduction will be done, as usual [8-12], in the close to threshold regime $r = r_0 + \varepsilon^2 r_2$, being $0 < \varepsilon \ll 1$ an auxiliary smallness parameter. We consider $\Delta = O(\varepsilon^0)$, where a complex Ginzburg-Landau (CGL) equation describes the laser in the absence of injection [10,11]. For smaller detunings, say $\Delta = O(\varepsilon)$, the nature of the bifurcation changes and a complex Swift-Hohenberg equation is obtained [12]. As for the injected signal we consider $F = O(\varepsilon^3)$ and $\omega, \theta = O(\varepsilon^2)$, which correspond to a weak field almost resonant with the free running laser emission frequency. Finally, we introduce slow spatial scales $(X = \varepsilon x, Y = \varepsilon y)$, and multiple slow time scales $(T_k = \varepsilon^k t, k = 1, 2, \ldots)$, and assume that the laser variables admit the expansion $G(x,y,t) = \sum_{n=1}^{\infty} \varepsilon^n G_n(X,Y,T_1,T_2,\ldots)$, $G = E, P, N$ [10,11]. Upon substituting all previous scalings into Eqs. (1)–(3) and equating equal powers in ε, an infinite hierarchy of equations is obtained which must be solved up to $O(\varepsilon^3)$. As usual, at this order a solvability condition must be imposed (in order to avoid unbounded solutions) which yields the searched equation. By undoing all the scalings the equation can be written as

$$s \partial_t u = i\theta u + \sigma(1-i\Delta)^{-1}(r - r_0 - |u|^2)u + i\nabla^2 u + E_0 \cos(\omega t), \qquad (5)$$

where $s = 1 + \sigma(1+i\Delta)/(1-i\Delta)$, and $u(\mathbf{r},t) = \exp(i\theta t)E(\mathbf{r},t)$. Equation (5) is a CGL equation with direct ac-driving which, for $E_0 = 0$, reduces to the laser CGL equation derived in [10]. Analytical, although very involved, information can be obtained from Eq. (5). Hence we consider for simplicity the case σ = 1 which, according to our numerics, is representative of many other cases. Now the CGL Eq. (5) can be written as

$$\partial_\tau u = (\mu + i\nu)u + (1+i\alpha)\nabla^2 u - |u|^2 u + f \cos(2\omega\tau), \qquad (6)$$

where $\tau = t/2$, $\mu = (r - r_0)$, $\nu = 2\theta$, $\alpha = \Delta^{-1}$, $f = (1-i\Delta)E_0$, and space has been normalized to $\sqrt{\Delta}$. In the following we choose, without loss of generality, *f* as a positive real number since the phase of *F* has not been fixed.

### 4. Ising and Bloch domain walls. Pattern formation

Let us consider now the limit of "strong" and "fast" modulation $f, \omega \gg 1$ in Eq. (6). By no means this limit is essential, but it allows a clear understanding of the role played by the bichromatic injection. In fact, the numerics we show in Sec. 5 do not verify that limit, eventhough they are qualitatively described by the features we show next.

In the limit $f = \eta^{-1}F$, $\omega = \eta^{-1}\Omega/2$, being $0 < \eta \ll 1$ a smallness parameter, we search for solutions to Eq. (6) in the form $u(\mathbf{r},\tau) = u_0(\mathbf{r},T,\tau) + O(\eta)$, where $T = \eta^{-1}\tau$ is a fast time scale. At $O(\eta^{-1})$ the solution reads $u_0(\mathbf{r},T,\tau) = (F/\Omega)\sin(\Omega T) + iU(\mathbf{r},\tau)$. The evolution equation for *U* is found as a solvability condition, which must be imposed at order $\eta^0$. The searched equation reads:

$$\partial_\tau U = [(\mu - 2\gamma) + i\nu]U + (1+i\alpha)\nabla^2 U - |U|^2 U + \gamma U^*, \qquad (7)$$

where $\gamma = \frac{1}{2}(F/\Omega)^2 = \frac{1}{8}(f/\omega)^2$. A main effect caused by the bichromatic injection is represented by the last, phase symmetry breaking term, which reduces the original phase invariance of the laser down to the discrete one $U \to -U$ (equivalent states whose phases differ by π). This allows the existence of *stable* domain wall solutions which, for $\alpha = \nu = 0$, are analytical in one spatial dimension, and are known as Ising (*I*) and Bloch (*B*) walls [10]:

$$U_I(x) = \pm\sqrt{\mu - \gamma} \tanh(\sqrt{\mu - \gamma}\, x/\sqrt{2}),$$
$$U_B(x) = \pm\left[\sqrt{\mu - \gamma} \tanh(\sqrt{2\gamma}\, x) \pm i\sqrt{\mu - 5\gamma}\,\text{sech}(\sqrt{2\gamma}\, x)\right]. \qquad (8)$$

The Ising wall (a dark line where $U = 0$) is stable for $\mu/5 < \gamma < \mu$, and the Bloch wall (a gray line) is stable for $0 < \gamma < \mu/5$. At $\gamma = \mu/5$ an Ising-Bloch transition occurs. Regarding the pattern forming properties of the system, a linear sability analysis of the spatially homogeneous solutions of Eq. (7) reveals that extended patterns arise, whenever $\alpha, \nu > 0$, with wavenumber *k* given by $\alpha k^2 = \nu - \gamma/\sqrt{1+\alpha^2}$ or $\alpha k^2 = \nu - \sqrt{(\mu - 2\gamma)^2 + \nu^2}/\sqrt{1+\alpha^2}$, depending on whether the trivial or the nontrivial (phase-locked) solutions is considered.

## 5. Numerics: Phase domains, labyrinths, and dark-ring cavity solitons

All previous analytical predictions were tested against the numerical integration of Eq. (6), and also of the full Maxwell-Bloch equations (1)–(3). A split-step method was implemented on a two-dimensional grid of 128×128 points with periodic boundary conditions. Numerics started from a spatially random distribution of the electric field. In the absence of injection vortex ensembles spontaneously developed, Fig. (a). For $\nu \leq 0$ Ising and Bloch domain walls were found as transient states, as usual in two-dimensional systems, Fig. (b). Also extended, labyrinthine patterns, Fig. (c), were found for $\alpha, \nu > 0$. Finally, for intermediate detuning values dark-ring CSs were found. Fig. (d) shows a transient state with two stable CSs and two contracting domains. In the final state four CSs are obtained.

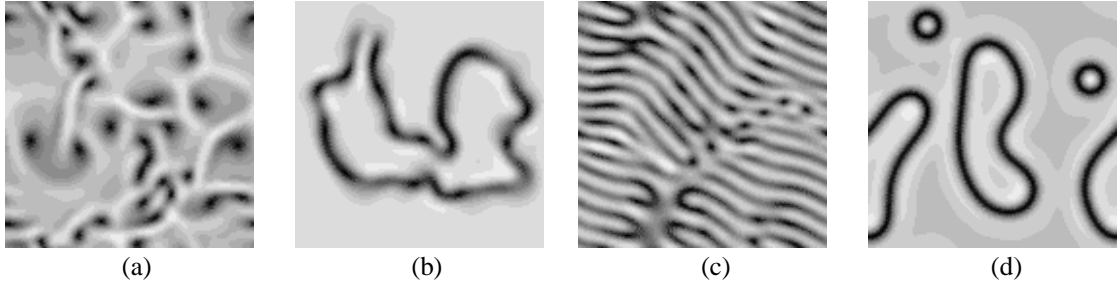

(a)　　　　　　　(b)　　　　　　　(c)　　　　　　　(d)

Figure. Laser intensity distribution according to the ac-driven complex Ginzburg-Landau equation (6) for $\mu = 1$ (set by normalization of), $\omega = 2\pi$, and $\alpha = 10$. Rest of parameters are $f = 0$ (a); $f = 7$, $\nu = 0$ (b); $f = 7.5$, $\nu = 0.6$ (c), $f = 10$, $\nu = 0.27$ (d).